\documentclass[osajnl, twocolumn, showpacs, superscriptaddress, 10pt]{revtex4-1} 
\usepackage{amsmath,amssymb,graphicx}
\usepackage{bbm}
\usepackage[colorlinks=true, citecolor=Red, urlcolor=Orange, linkcolor=blue]{hyperref}
\usepackage[usenames, dvipsnames]{color}
\begin{document}

\title{Residual Amplitude Modulation in Interferometric Gravitational Wave Detectors }

\author{Keiko Kokeyama}\email{Corresponding author: keiko@lsu.edu}
\affiliation{Louisiana State University, Physics and Astronomy, 202 Nicholson Hall, Baton Rouge, 70806, LA, USA}

\author{Kiwamu Izumi}
\affiliation{University of Tokyo, Department of Astronomy, 7-3-1 Hongo Bunkyo-ku, Tokyo, 113-0033, Japan}
\affiliation{LIGO Hanford Observatory, PO Box 159 Richland, WA 99352, USA}
\author{William Z. Korth}
\affiliation{California Institute of Technology, 1200 E. California Blvd. Pasadena, CA 91125, USA}
\author{Nicolas Smith-Lefebvre}
\affiliation{California Institute of Technology, 1200 E. California Blvd. Pasadena, CA 91125, USA}
\author{Koji Arai}
\affiliation{California Institute of Technology, 1200 E. California Blvd. Pasadena, CA 91125, USA}
\author{Rana X. Adhikari}
\affiliation{California Institute of Technology, 1200 E. California Blvd. Pasadena, CA 91125, USA}

\begin{abstract}
The effects of residual amplitude modulation (RAM) in laser interferometers using heterodyne sensing
can be substantial and difficult to mitigate. In this work, we analyze the effects of RAM on a complex laser
interferometer used for gravitational wave detection.
The RAM introduces unwanted offsets in the cavity length signals
and thereby shifts the operating point of the optical cavities from the nominal point via feedback control.
This shift causes variations in the sensing matrix,
and leads to degradation in the performance of the precision noise subtraction scheme
of the multiple-degree-of-freedom control system.
In addition, such detuned optical cavities produce an
opto-mechanical spring, which also varies the sensing matrix.
We use our simulations to derive requirements on RAM  for the Advanced LIGO detectors,
and show that the RAM expected in Advanced LIGO will not limit its sensitivity. 
\end{abstract}


\maketitle 

\section{Introduction}
\label{sec:intro}
The direct detection of gravitational waves (GW) is a challenging but
significant goal for fundamental physics and astronomy in the near future.
There are several laser interferometric detector projects around the globe that aim to directly detect GWs
from astrophysical sources, such as Advanced LIGO (aLIGO)~\cite{PF:RPP2009, aLIGO}, 
Advanced VIRGO~\cite{2006virgo}, GEO-HF~\cite{2006GEO}, and KAGRA~\cite{KAGRA}.

These interferometers are kilometer-scale Michelson interferometers with coupled Fabry-Perot
cavities used to enhance the sensitivity to the spacetime strain induced by the GWs. The expected
strains from astrophysical events should produce displacements of order $10^{-19} $ m in these
interferometers.

To measure such a small difference with a high signal-to-noise ratio,
the lengths of the optical cavities in the interferometer
must be controlled by precise positioning of all the constituent optics, ensuring that the 
optical response to the GW is linear.

The interferometric length signals are derived by heterodyne detection using a variant of
the Pound-Drever-Hall (PDH) cavity locking technique\,\cite{Sigg:Readout}:
the laser light is phase-modulated at a certain frequency before being injected
into the interferometer. A signal is then generated by the beat
between the fields at different frequencies, such as the carrier field and modulation sidebands.
In the aLIGO interferometer, phase modulation (PM) at two different frequencies
is introduced to robustly extract the length signals for the multiple degrees of freedom (DoFs).
In practice, the obtained heterodyne signals induce
the cross-couplings between the GW DoF and auxiliary DoFs.

Due to imperfections (explained in later sections),
the electro-optic modulator (EOM) used to impose PM sidebands on the optical field also introduces 
some residual amplitude modulation (RAM).
This RAM is at the same frequencies
as the PM and may therefore introduce spurious signals and offsets
when the laser fields are sensed by photodetectors and demodulated
to generate error signals.
This may lead to changes in the frequency response to GWs (i.e. the calibration) and
also to spurious couplings of laser noise to the GW readout channel.

In this paper, we present a model of how RAM affects
the response of the GW interferometer and impacts its sensitivity.
In Section~\ref{sec:RAMintro}, we introduce RAM and describe its effect on the GW detectors.
In Section~\ref{sec:setups} we give an overview of our simulation setup.
Section~\ref{sec:results} presents the results: the predicted effect of RAM effect on aLIGO sensitivity.
Simulation parameters and details of the sensing and control scheme used in our calculation can be 
found in Appendix~\ref{sec:app:param} and~\ref{sec:app:ISC}, respectively.

\section{RAM and advanced GW detectors}
\label{sec:RAMintro}
In this section, we first give an overview of the aLIGO optical configuration
and its sensing and control scheme.
Then, we introduce the issue of RAM and show how it may lead to problems with sensing and control.

\begin{figure*}[t]
\begin{center}
\includegraphics[width=1.5\columnwidth]{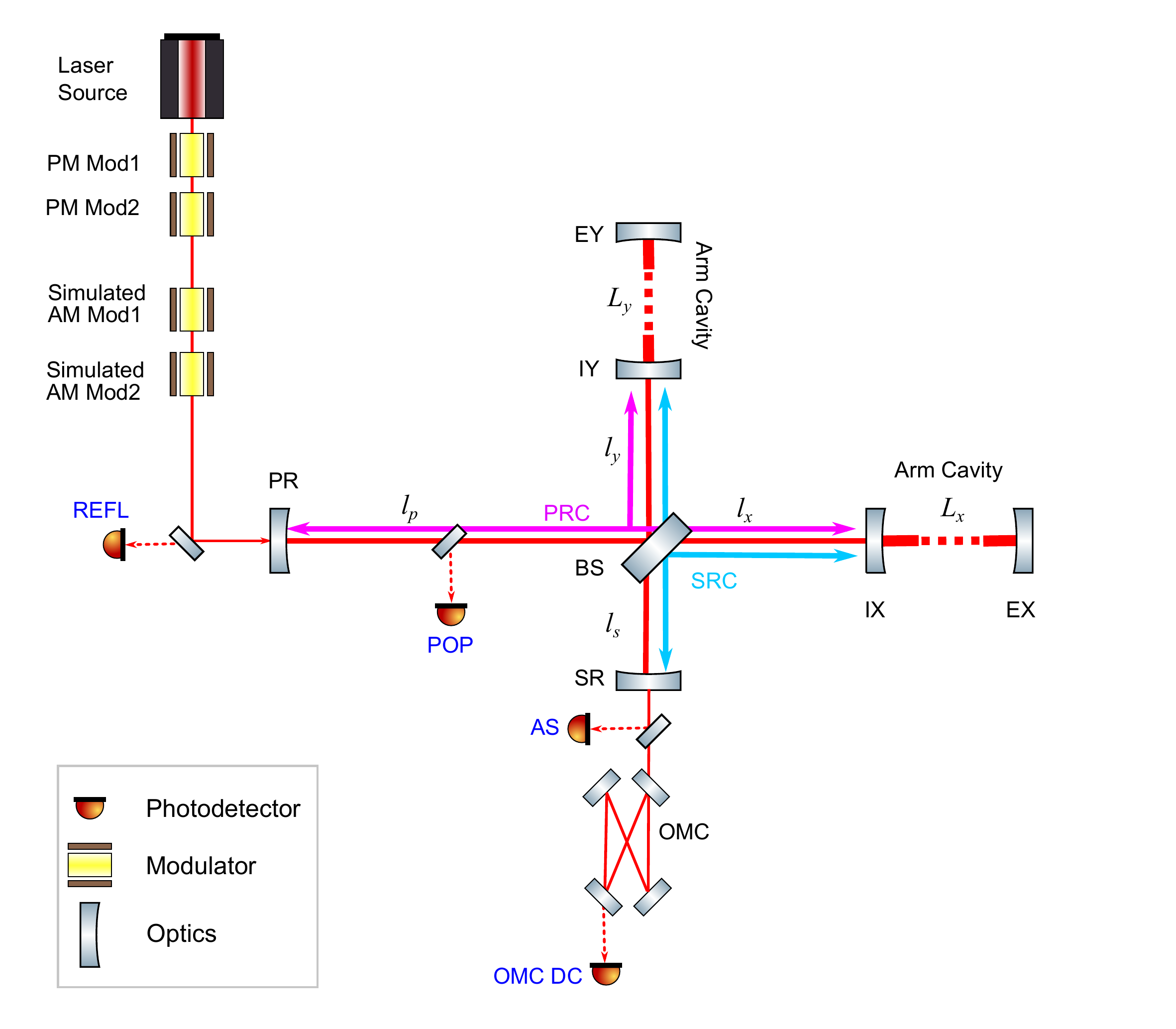}
\caption{Full interferometer optical configuration. PR: power recycling mirror,
PRC: power recycling cavity,
SR: signal recycling mirror,
SRC: signal recycling cavity,
IX, IY: input test mass at X or Y arm, respectively,
EX, EY: end test mass at X or Y arm, respectively,
BS: beamsplitter,
REFL: reflection port, AS: anti symetric port, OMC:
output mode cleaner, PO: pick off port, OMC DC, DC readout port at the OMC transmission,
PM Mod1 and 2: phase modulators for sideband 1 and 2, respectively
AM Mod1 and 2: amplitude modulators to simulate the residual amplitude modulation.
$l_p$: PR to BS, $l_s$: SR to BS, $l_x {\rm and} l_y$: BS to IX and IY, respectively,
$L_x$ and $L_y$: x and y arm, respectively.}
\label{fig:config}
\end{center}
\end{figure*}

\begin{table}[t]
\begin{center}
\begin{tabular}{cc}
\hline
           & Definition                       \\ \hline
DARM& $(L_x - L_y)/2$                      \\ \hline
CARM & $(L_x + L_y)/2$                     \\\hline
MICH &   $l_x - l_y$ \\ \hline 
PRCL &  $l_p + (l_x + l_y)/2 $\\ \hline
SRCL &  $ l_s + (l_x + l_y)/2$ \\ \hline
\end{tabular}
\caption{The definitions of the five length degrees of freedom to control.}
\label{tb:5dof}
\end{center}
\end{table}

\subsection{Optical configuration and sensing scheme of aLIGO}
Fig.~\ref{fig:config} shows the optical configuration of aLIGO.
Each arm of the interferometer is a 4-km Fabry-Perot cavity that enhances
the instrument's response to GW signals. A power recycling cavity (PRC) enhances the effective incident laser power,
and a signal recycling cavity (SRC) tunes the detection bandwidth.
There are five DoFs to control: DARM, the differential motion of the two arm cavities
(the GW channel);
CARM, the common motion of the arm cavities;
MICH, the difference between two short arms of the Michelson interferometer;
PRCL, the PRC length, i.e.,
the average distance between the power-recycling mirror (PRM)
and the two input test masses (ITMs, IX and IY in Fig.~\ref{fig:config});
SRCL, the SRC length, i.e., the average distance
between the signal recycling mirror (SRM) to the two ITMs.
The explicit definitions of these DoFs are summarized in Table~\ref{tb:5dof}.

To robustly extract length signals for the five DoFs,
two sets of PM sidebands are used.
Mirror motion induces phase modulation of the carrier and sideband laser fields which experience different resonant conditions in the interferometer due to the careful choice of modulation frequencies. Demodulated signals from photodetectors that sample fields extracted from the interferometer produce signals that contain information about the various DoFs.


\subsection{Electro-optic modulator and RAM source}
\begin{figure}[t]
\centerline{\includegraphics[width=\columnwidth]{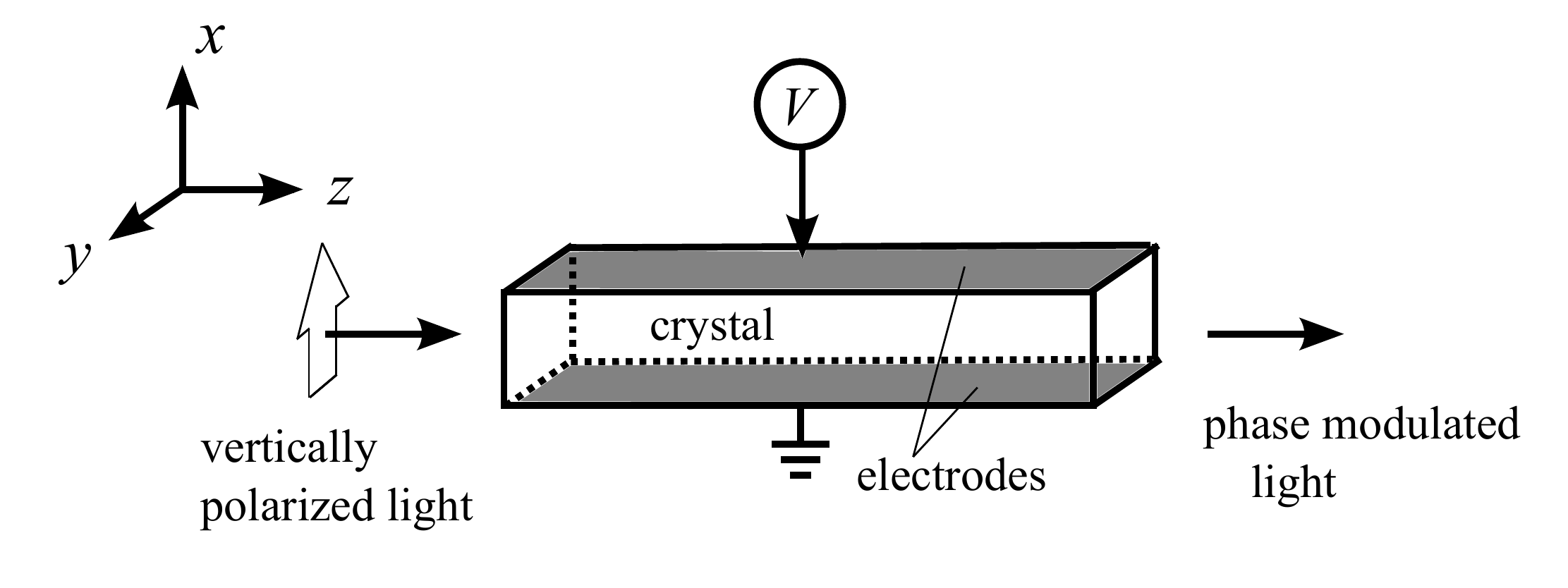}}
\caption{Electro-optic modulator. The modulation volage $V$ is applied along the crystal
in $x$ axis, which is one of the crystal principal axes.
The polarized field in $x$ axis is phase-modulated
after passing through the crystal. The beam propagates in $z$ direction.}
\label{fig:eom}
\end{figure}

An electro-optic modulator (EOM) is a device used to modulate the phase of a laser field.
An EOM consists of a Pockels cell---a crystal exhibiting a birefringence that depends linearly 
on the applied electric field---and a set of electrodes.
The electrodes, attached on the top and bottom surfaces of the crystal, apply a voltage
along one of the principal axes of the crystal.
Fig. \ref{fig:eom} shows the EOM crystal with voltage applied along $x$ axis,
and the laser light polarized along the $x$ axis as an input.
As the laser light passes through the crystal, the birefringence of the crystal changes due to the
externally applied electric field and the laser light is phase-shifted.
The phase shift induced by the electric field is written as
$\triangle \phi = \pi n_x ^3 r V/ \lambda$ where $n_x,~ r, ~V, ~\lambda$
are the unperturbed refractive index in the $x$ direction, an electro-optic coefficient,
the applied voltage, and the laser wavelength, respectively~\cite{crystal}.

During the phase modulation process,
unwanted amplitude modulation is also imposed due to
the following effects~\cite{Whittaker, WongHall, etalon} which are difficult to avoid in practice:
(1) \emph{Axis mismatch between the incident polarization and the crystal orientation}.
When the input beam polarization axis does not align with one of the crystal axes
and the axis of the applied electric field,
the projections of the input field onto the crystal's orthogonal axes obtain different phase shifts.
For example, in Fig.~\ref{fig:eom}, when the beam polarization is not along the $x$ direction,
the laser field polarized in the $x$ direction obtains a phase shift,
while the laser field polarized in the $y$ direction obtains a different phase shift. This leads to an 
effective rotation of the transmitted beam's polarization, which, upon its subsequent interaction 
with polarizing optics, leads to AM.
(2) \emph{Etalons in the crystal}.
Due to the finite reflectivities of the crystal faces, some light is circulated within the crystal. 
The multi-pass light field experiences a frequency-dependent phase shift and amplitude 
envelope due to this etalon effect.

In practice, the use of good anti-reflection (AR) coatings and wedged crystal faces 
makes (1) the dominant contribution to RAM.

Whatever the coupling, temperature fluctuations are usually the major driving force of RAM generation.
Temperature drifts affect the mounting orientation of the crystal
through the thermal expansion of the mounting material,
and also lead directly to expansion of the crystal and modulation of its static birefringence.
As such, a first approach to reducing RAM is simple temperature 
stabilization of the crystal and enclosure. More sophisticated techniques involve measuring the RAM 
optically and actively feeding back to either the modulation voltage~\cite{WongHall} or the crystal 
temperature~\cite{Li}.

%

\subsection{Residual Amplitude Modulation}
Residual amplitude modulation (RAM) is  imposed at the same frequency
as that of the intentional phase modulation.
A laser field whose phase and amplitude are modulated at a frequency $\omega_m$
can be expressed as
	\begin{eqnarray}
\begin{split}
		 E = E_\mathrm{in} \left[ 1+\Gamma_a \sin( \omega _m t+ \phi) \right] \\
		\times \exp \left[ i\omega _0 t
		+ i\Gamma _{p} \sin \omega _{m} t \right],
\end{split}
	\end{eqnarray}	
where $E_\mathrm{in}$ is the amplitude of the field, $\omega_0$ denotes the angular
frequency of the carrier field,
$\Gamma_a$ and $\Gamma_p$ are the modulation depths of the RAM and PM respectively,
and $\phi$ is the relative phase between the RAM and PM terms.
In order to evaluate the amount of the RAM relative to the PM,
we define the RAM-to-PM ratio as
	\begin{equation}
		\eta \equiv \frac{\Gamma_a}{\Gamma_p}.
	\end{equation}
When PM is imposed at two different frequencies, the field with RAM can be written as,
	\begin{eqnarray}
\begin{split}
E = E_\mathrm{in} \bigl[ 1 + \Gamma_{a1} \sin( \omega _{m1} t+ \phi_{\rm A1}) 
		+\Gamma_{a2} \sin( \omega _{m2} t+ \phi_{\rm A2}) \bigr] \\ 
\times \exp \bigl[ i\omega _0 t 
		 + \Gamma _{p1} \sin (\omega_{m1} t + \phi _{\rm P1}  \qquad \qquad \\
		+ \Gamma _{p2} \sin (\omega _{m2} t + \phi _{\rm P2}) \bigr]. \qquad 
\end{split}
	\end{eqnarray}		
where $\phi_{\rm A1}$, $\phi_{\rm A2}$, $\phi_{\rm P1}$ and $\phi_{\rm P1}$
are the arbitrary phases of the RAM and phase modulation terms.

Since they are at the same frequencies as the PM sidebands, the RAM sidebands introduce unwanted signal offsets
when interferometer signals for a given DoF are extracted from the beat between
the carrier field and a control sideband on a photodetector.
Because the interferometer optics' positions are controlled
by servoing these error signals to zero, the offsets due to RAM can change their positions, affecting 
the interferometer response and degrading its sensitivity to GWs.

\section{Simulation}
\label{sec:setups}

To evaluate the effect of RAM on the GW sensitivity,
we used a frequency-domain simulation tool, Optickle~\cite{optickle, optickle2}.
The aLIGO optical configuration (shown in Fig.~\ref{fig:config})
was modeled with parameters summarized in Appendix~\ref{sec:app:param}.

There are three signal extraction ports in the interferometer:
reflection (REFL) port, anti-symmetric (AS) port, and
pick-off port (POP, pick off field from PRC).
An input matrix (not shown) links the DoFs in the left column of 
Table~\ref{tb:inout} to the sensing signals in the middle column.
We assigned these sensing ports
so that each port has the maximum sensitivity to the assigned DoF.
Similarly, an output matrix feeds the derived control signals
to an optic position actuation path in the right column.

The gravitational wave channel (DARM) is detected at the output mode cleaner
transmission port (OMC DC).
This ``DC readout'' is a homodyne detection method,
which requires the introduction of a small DARM offset to bring the AS port slightly off of a dark 
fringe (producing linear sensitivity to the DARM DoF).
This scheme is proposed for the GW readout
due to its better signal-to-noise ratio compared with
the heterodyne readout~\cite{DCreadout, DCreadout2, DCreadout3}.

The actuation mapping on the right-hand side of Table~\ref{tb:inout} is chosen so as to maintain the 
orthogonality of the DoFs as best as possible; in practice, there is always some cross-coupling 
between DoFs. In the case of aLIGO, DARM is polluted most seriously by MICH and SRCL. 
To mitigate this effect, feedforward is applied from these DoFs to DARM---this feature is included in our simulation.
See Appendix~\ref{sec:app:ISC} for more information about the sensing and control scheme.

The frequency dependence from force to displacement
of the suspended test masses is approximated using a transfer function where
the response is flat up to 1 Hz peak and then drops as $f^{-2}$.
This is an approximation of the transfer function of quadruple suspensions
which suspend the main optics in aLIGO~\cite{QUAD}.


\begin{table}[t]
\begin{center}
\begin{tabular}{ccc}
\hline
DoF & Input & Output\\ \hline \hline
 DARM& OMC DC& EX -- EY \\ \hline
CARM & REFL f1 & EX + EY\\ \hline
MICH & REFL f2 Q& $\sqrt 2 $ BS --PR +SR \\ \hline
PRCL& REFL f2 I& PR \\ \hline
SRCL & POP f2& SR\\ \hline
\end{tabular}
\caption{Input-output chain for aLIGO setup.\
The input column shows the sensors used to detect each DoF
with I (I-phase) or Q (Q-phase) demodulation phase.
f1 and f2 represent the demodulation frequency.
The output column shows which optics are fed back to
for control of each DoF.}
\label{tb:inout}
\end{center}
\end{table}

\section{Simulation results}
\label{sec:results}

\subsection{Operating point offsets}

\begin{figure}[]
\begin{center}
\vspace{0mm}
\includegraphics[width=\columnwidth]{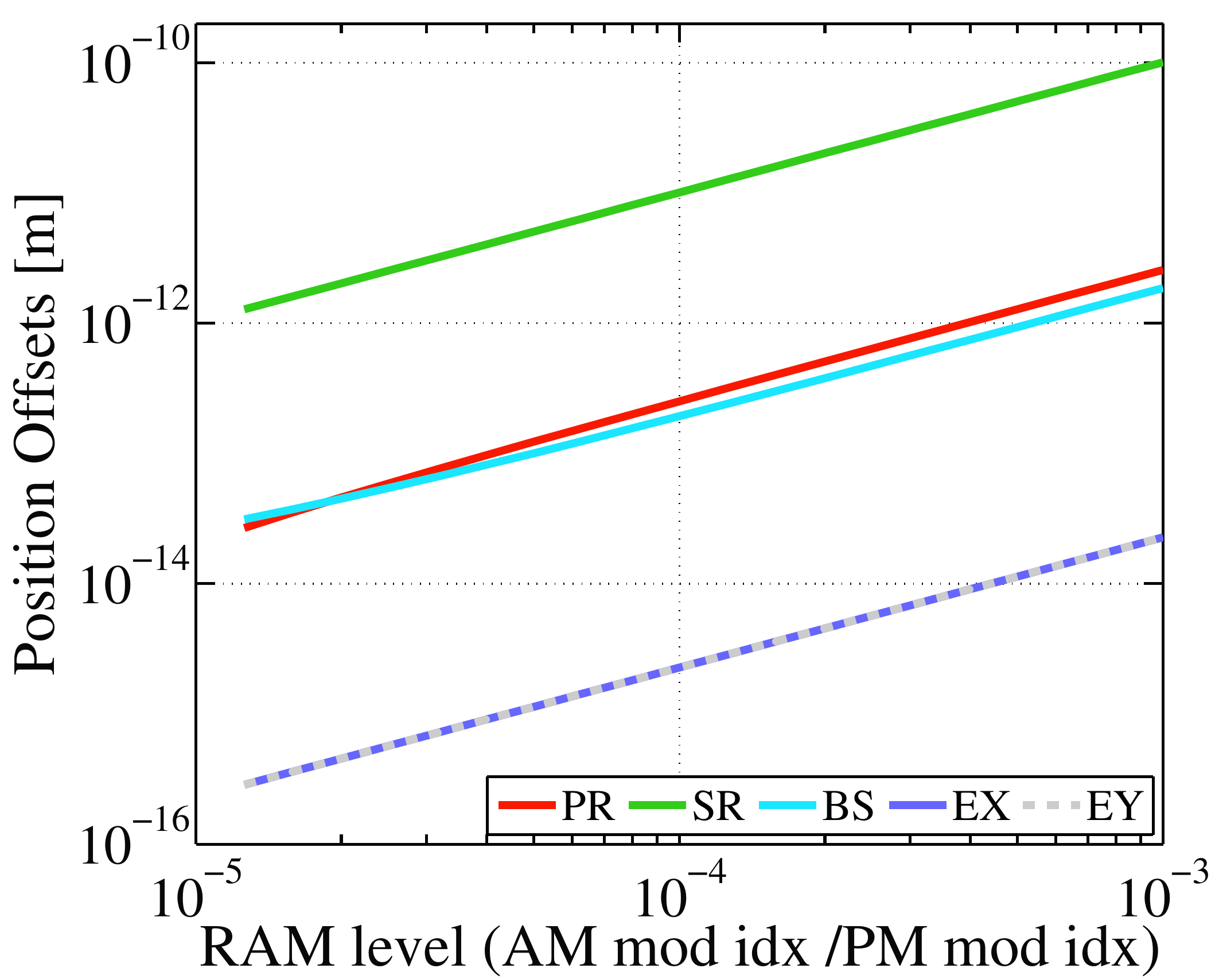}
\caption{The length offsets on each optic to make the length error signal
for each DoF zero, as a function of the RAM level.
Larger RAM adds more offsets on the error signals,
resulting in the length offsets of the optics.}
\label{fig:offset}
\end{center}
\end{figure}

The length offsets of each DoF due to RAM are shown in Fig.~\ref{fig:offset}.
These was calculated by the following iterative process: 
First, the error signal offset due to RAM is calculated.
Then, the operating point of each DoF is changed
so that the error signals are zero for the corresponding DoF.
The error signal now has the different RAM offsets
because the interferometer response is slightly changed by the length offsets.
Therefore, a new length offset is added to make the new error signal zero.
We iterate this process until the RAM offset and the operation point converge to below $10^{-12}$ times 
the laser wavelength.

The largest effect of RAM is on the SRCL DoF, with smaller effects seen on PRCL and MICH.
We can also see that a larger RAM level produces larger length offsets.
CARM and DARM (EX and EY in Fig.~\ref{fig:offset}) are less sensitive to RAM
because the frequency discriminants of the corresponding error signals are enhanced by the Fabry-Perot 
arm cavities. In addition, the DARM DoF is read out using homodyne detection (i.e., there are no RF sidebands) 
and is therefore only sensitive to RAM via the cross-coupling effects described above.

\subsection{Effects on opto-mechanical response}
\begin{figure}[t]
    \centering
\includegraphics[width=\columnwidth]{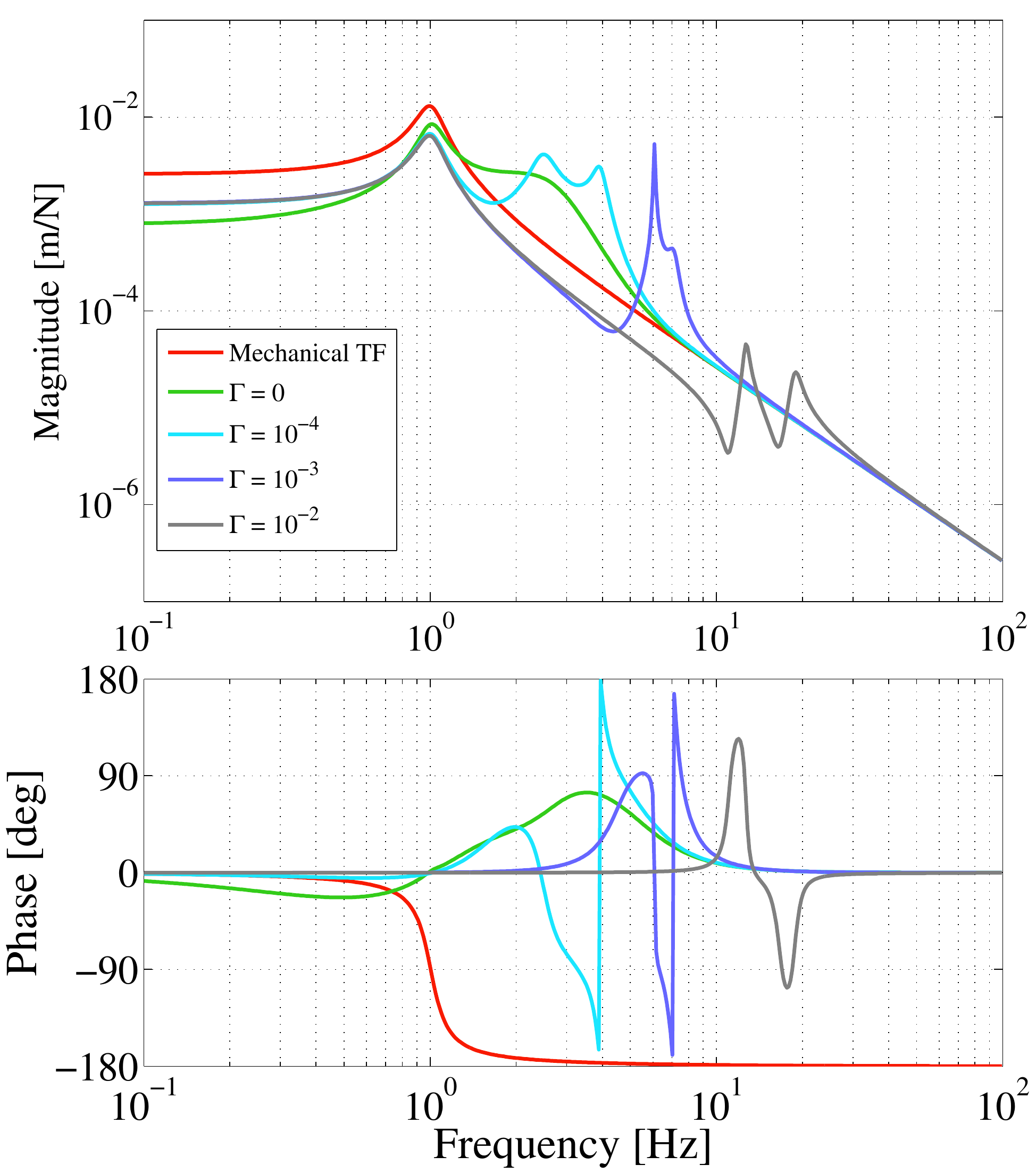}
    \caption{(color online) Mechanical transfer function of the suspension in [m/N] (red)
      and the modified opt-mechanical transfer functions with various RAM-to-PM ratios and 
      DARM offset for DC readout. Length offsets produce detuning of the interferometer's 
      coupled optical cavities. In particular, the SRC---which experiences the strongest 
      RAM-induced effect---is pushed 0.078 degrees off resonance when $\Gamma =10^{-3}$, 
      and even this small a detuning is sufficient to create a potentially unstable opt-mechanical spring.}
    \label{fig:RAMspring}
\end{figure}

The opto-mechanical response is altered when RAM is present.
Fig.~\ref{fig:RAMspring} shows this transfer function from force to displacement of test masses,
with a PM-to-RAM ratio ranging from  $\Gamma =$ 0 to $\Gamma = 10^{-2}$.
Opto-mechanical peaks appear even
even when RAM is absent, $\Gamma =0$, because of the slight DARM offset for DC readout
(3\,pm is used here).
With $\Gamma \gtrsim 10^{-4}$ or more, the opt-mechanical peaks become sharper
due to the length offsets shown in the previous subsection.
These small offsets---particularly that to SRCL---result in detuned operation
of the signal recycling cavity (0.078 degrees for $\Gamma = 10^{-3}$), which gives
rise to the opto-mechanical spring.
Such resonances may add instability to the control loop and should be avoided.

\subsection{Strain sensitivity and loop noise}
\begin{figure}[h]
\centering
\includegraphics[width=\columnwidth]{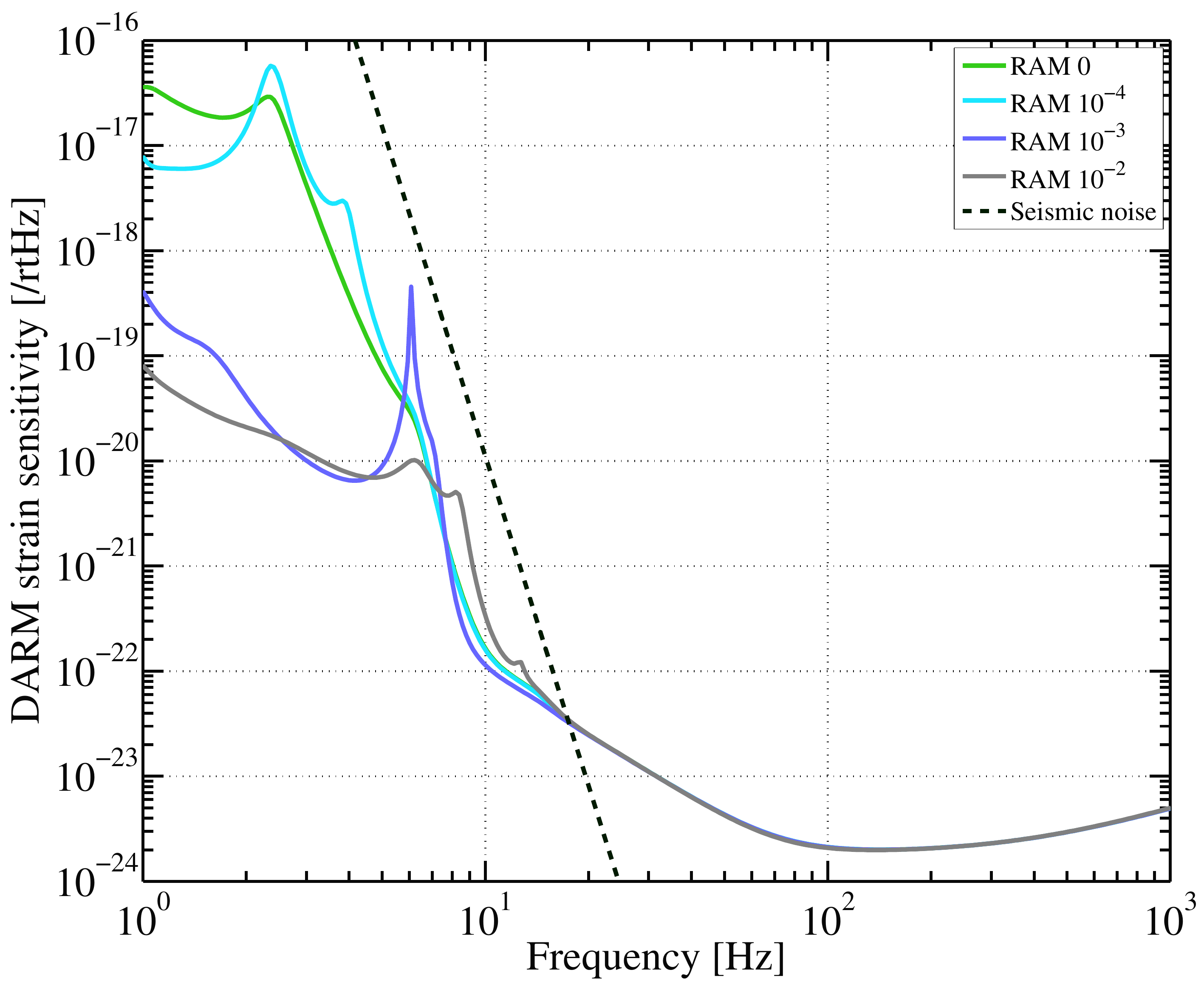}
    \caption{(color online) The quantum-noise-limited DARM sensitivity with
      $\Gamma = 0, 10^{-4}, 10^{-3},$ and $10^{-2}$.
      The intrinsic quantum noise of DARM and the quantum noise contributions
      (i.e., the noise at each sensor propagated through to each control signal)
      from auxiliary DoFs  are considered.
      The sensitivity at low-frequencies is limited by the cross-coupled noise from auxiliary DoFs
      (see, Fig. \ref{fig:DARMloop} for the brakedown of the cross-couplings).
      The cross-coupled noise shape was changed due to the length offsets generated by RAM.
      The gray line is the estimated seismic noise level,
      which is predicted to limit the GW sensitivity of aLIGO in this low-frequency region.}
  \label{fig:result}
\end{figure}

Due to this change in optical response, the resulting strain sensitivity is altered
as shown in Fig.~\ref{fig:result}.
There is no significant difference with $\Gamma \approx 10^{-4}$, the level measured in the EOM in use at the LIGO Livingston Observatory.
Even with larger Gamma, the quantum noise is little affected above 10 Hz.

\begin{figure}[h]
  \centering
  \includegraphics[width=\columnwidth]{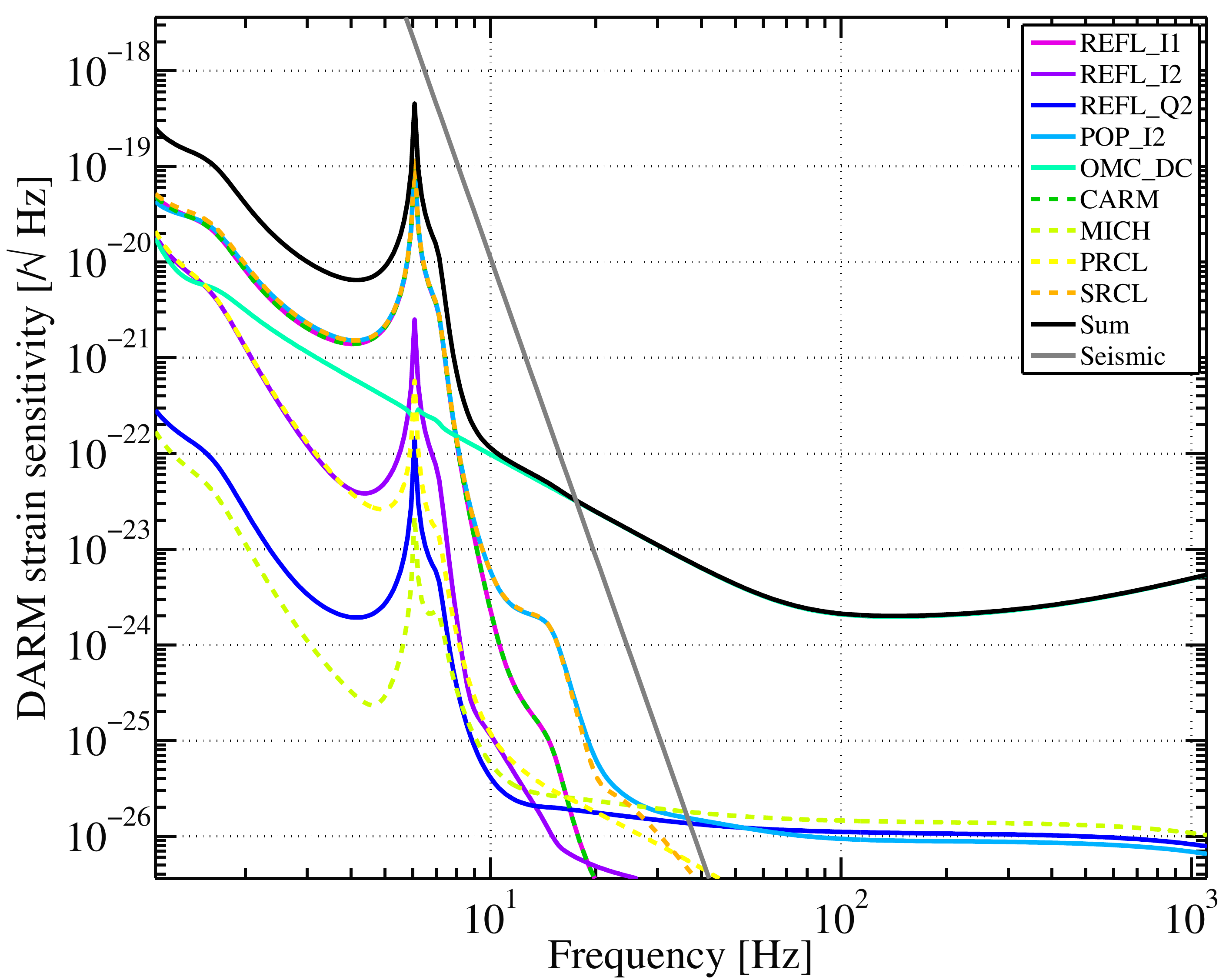}
\caption{(color online) Loop noise picture of DARM  when $\Gamma = 10^{-3}$.
  Shown here is the DARM sensing noise (OMC\_DC),
  as well as the the cross-coupled quantum noise (solid lines)
  and control noise (dashed lines) from each auxiliary DoF.
  The opt-mechanical spring appears through the loop noise mixing
  while the OMC\_DC noise floor is slightly changed by RAM.
  The low frequency interferometer performance is limited mainly by 
  the SRCL control noise and SRCL sensor (POP\_I2) in our control loop model.}
     \label{fig:DARMloop}
\end{figure}

Note that quantum noise contributions from auxiliary DoFs are included in Fig.~\ref{fig:result}, in addition to the 
quantum noise intrinsic to DARM. In Fig.~\ref{fig:DARMloop}, we see the breakdown of the DARM sensitivity 
for the case of $\Gamma = 10^{-3}$. The noise intrinsic to the DARM loop is shown in the light green 
trace (``OMC\_DC''), and all other traces are cross-couplings. Below around 12\,Hz, the total noise 
(i.e., the red trace in Fig.~\ref{fig:DARMloop}) is dominated by contributions from other DoFs. This is 
the reason for the seeming enhancement in DARM sensitivity with increasing RAM; the contribution 
from cross-coupling is indeed suppressed due to the modified response, but the noise intrinsic to the 
DARM loop experiences no such reduction. Therefore, the ideal quantum-noise-limited GW sensitivity 
is not improved by the introduction of RAM. To the contrary, it is diminished for higher levels of RAM, and 
it is not effected appreciably by the levels considered here. In any case, the sensitivity of aLIGO to GWs 
is predicted to be completely dominated by seismic noise at these low frequencies, as shown by the 
gray trace in Fig.~\ref{fig:result}.


\appendix
\section{Simulation parameters}
\label{sec:app:param}
The optical parameters are chosen to model the aLIGO Livingston interferometer, as listed in
Table~\ref{tb:param} and found in \cite{aLIGOparam, aLIGOparam2}.
The servo filters are chosen
to have a unity gain frequency (UGF) of 200\,Hz for DARM,
50\,kHz for CARM, and 20\,Hz for PRCL, 15\,Hz for MICH and 10\,Hz for SRCL.

\newcommand{\1}{\mbox{1}\hspace{-0.5em}\mbox{l}}

\begin{table}[t]
\centering
\begin{tabular}{cc}
\hline
Arm lengths &   3994.5 m       \\\hline
Arm cavity FWHM / FSR  &  84.2344 Hz / 37.526 kHz\\ \hline
PRC length   &  57.6562 m\\ \hline
PRC FWHM / FSR  & 13.190 kHz / 2.5998 MHz \\ \hline
SRC length   &  56.0082 m\\ \hline
SRC FWHM / FSR  & 179.18 kHz / 3.7526 MHz  \\ \hline
Schnupp Asymmetry & 8 cm    \\ \hline
Short Michelson arm length & 5.3428 m  (average)\\ \hline
PRM transmissivity & 0.03 \\ \hline
SRM transmissivity &0.35 \\ \hline
ITM transmissivity & 0.14 \\ \hline
ETM transmissivity & 5 ppm\\ \hline
Input laser power & 135 W or 25 W\\ \hline
Optics loss (per mirror) & 30 ppm \\ \hline
OMC transmissivity & 0.99 \\ \hline
PM1 frequency & 9.099471 MHz \\ \hline
PM2 frequency & 45.497355 MHz \\ \hline
PM1 modulation index & 0.1 \\ \hline
PM2 modulation index & 0.1 \\ \hline
\end{tabular}
\caption{aLIGO parameters used in our simulation. FWHM; the full width at half maximum,
FSR; free spectral range.}
\label{tb:param}
\end{table}


\section{Sensing Scheme}
\label{sec:app:ISC}
In this section, we describe in detail the sensing signals
and the control scheme used in our simulation,
including a brief description of the multiple-DoF control method.

\subsection{Sensing and Control} 
\begin{figure*}[t]
\begin{center}
\includegraphics[width=2.2\columnwidth]{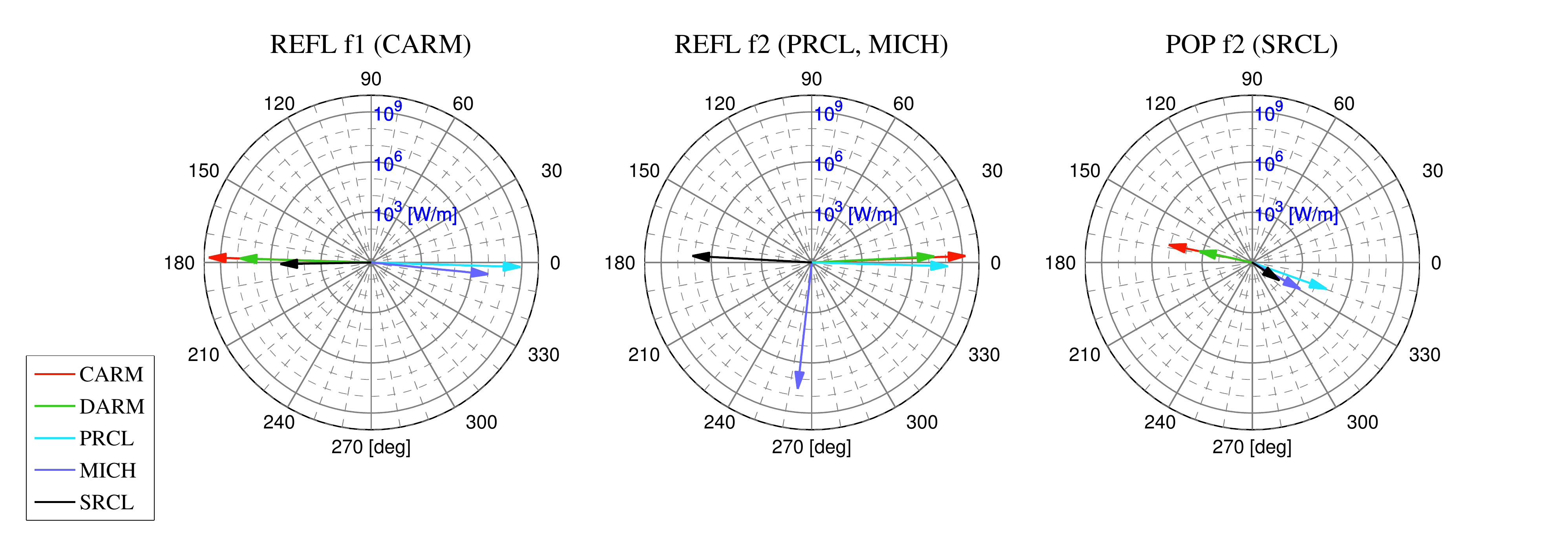}
\caption{Radar representation of the sensing matrix. The arrow lengths represent
the signal strength [W/m], and the angles show the demodulation phases
for each signal. OMC DC sensor for DARM is not shown because it is a DC readout signal
and does not have demodulation phase.}
\label{fig:radar}
\end{center}
\end{figure*}

Fig.~\ref{fig:radar} is the {\it radar plot} representing the signals in each sensors.
The five DoF signals in each sensor (i.e., in each radar) are shown as arrows.
The arrow length is the signal strength in W/m in logarithmic scale
and the angles are the optimum demodulation phases in degrees.
Because the DC readout signals (OMC DC sensor) for DARM is a DC signal
and does not have demodulation phase, it is not listed in the radar plot.

The desired signal at each sensor is obtained
by demodulating at the optimum demodulation phase for the targeted DoF.
Undesired signals mix into the obtained signal
with a ratio determined by the relative phase separation from the desired signal.
The sensors are chosen to, as much as possible, separate the various length signals from one another to maximize sensitivity to the desired DoF and minimize cross-coupling from other DoFs.
The signals shown in \ref{fig:radar} are calculated at 20 \,Hz, with an input power of 25\,W,
and a differential arm offset of 3\,pm.

Table~\ref{tb:sensmat} shows the conventional sensing matrix of our model.
This matrix is often used to express the signal sensitivity
to the different DoFs, at each detection port, at a certain frequency~\cite{sensmatrix}.
The first column of a sensing matrix shows the sensors, which is a choice
of the detection port, demodulation frequency, and demodulation phase (I- or Q-phase),
and the rows of a sensing matrix show the gains of the optical response for each DoF.
The diagonal elements of the matrix correspond to the sensitivity of the given sensor to the 
DoF to be controlled by it,
and the off-diagonal elements show the contributions from the other DoFs.
One sensor is fed back to one DoF for the control, therefore, we have five sensors
to control the five DoFs.

For example, in our model, 
OMC DC is chosen for the DARM control, having the maximum sensitivity to DARM,
with a some contribution from PRCL, MICH and SRCL.

\begin{table*}[ht]
\centering
\begin{tabular}{ccccccc}
\hline
 Sensor       &Phase & CARM & DARM & PRCL & MICH & SRCL                                         \\ \hline \hline
REFL I1 & 178$^{\circ}$ & $ 5.10\times 10^{9 }$ & $ 7.16\times 10^{7 }$ & $ -8.23\times 10^{8 }$ & $ -1.08\times 10^{7} $ & $ 2.67\times 10^{5 }$ \\ \hline
OMC DC & N/A & $ 6.63\times 10^{0} $ & $ 1.77\times 10^{5} $ & $ 1.14\times 10^{3 }$ & $ 6.32\times 10^{1} $ & $ 2.19\times 10^{2} $\\ \hline
REFL I2& -2$^{\circ}$& $ 1.54\times 10^{9} $ & $ 2.15\times 10^{7} $ & $ 1.36\times 10^{8} $ & $ -3.35\times 10^{6} $ & $ -1.39\times 10^{7} $\\ \hline 
REFL I2 & -96$^{\circ}$ & $ -2.37\times 10^{8} $ & $ -3.46\times 10^{6} $ & $ -1.15\times 10^{7 }$ & $ 3.97\times 10^{7 }$ & $ 7.85\times 10^{5} $\\ \hline
POP I2 & 34$^{\circ}$ & $ -1.18\times 10^{5} $ & $ -1.66\times 10^{3} $ & $ 4.92\times 10^{4} $ & $ 1.84\times 10^{3} $ & $ 8.50\times 10^{1 }$\\ \hline
\end{tabular}
\caption{Sensing matrix. DC readout offset (DARM DC offset) of 3\,pm is assumed.
They are optical gains [W/m] at 20\,Hz.  Demodulation phases are optimized for the diagonal elements in each sensor.}
\label{tb:sensmat}
\end{table*}

\subsection{Signal Mixture}

\begin{figure*}[ht]
\begin{center}
\includegraphics[width=1.5\columnwidth]{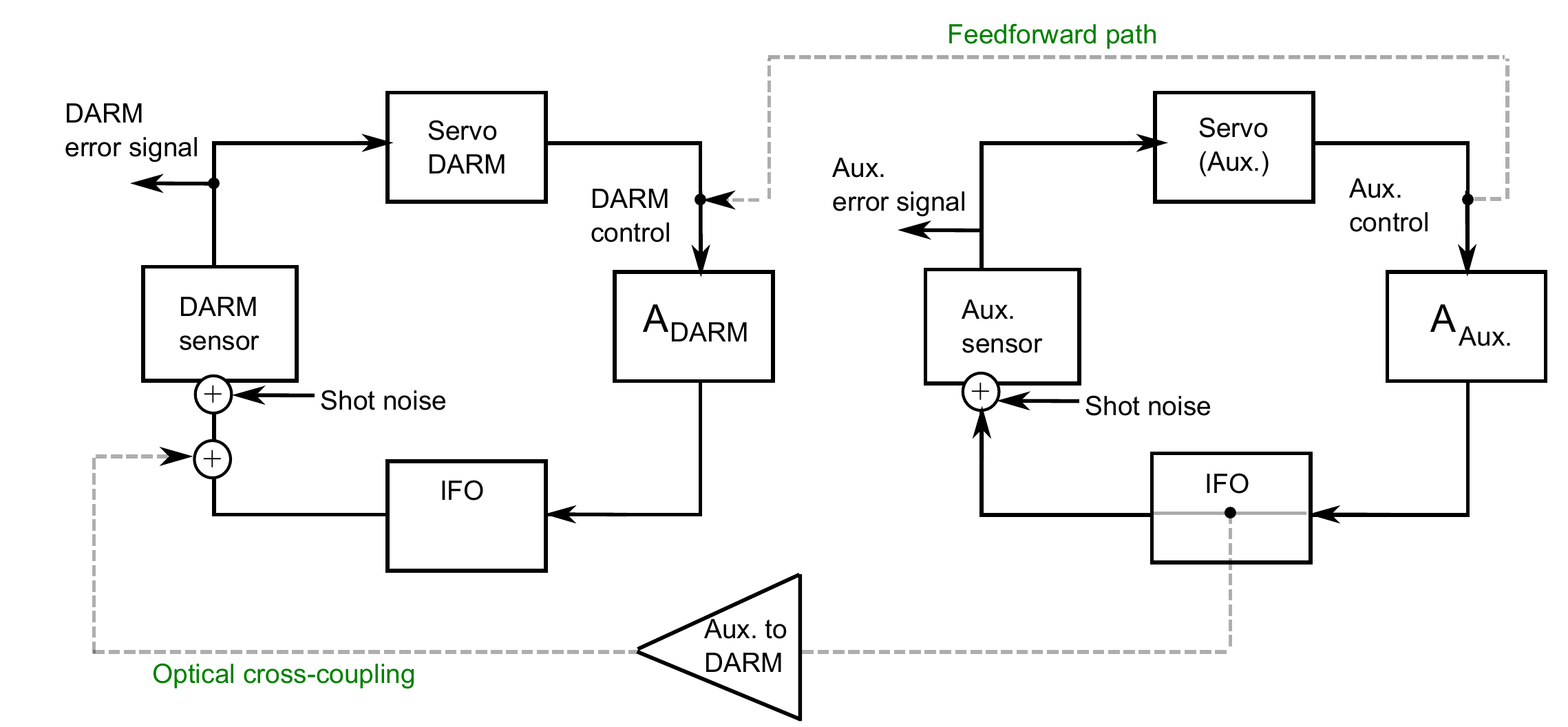}
\caption{Diagram of the signal mixture from an auxiliary to the DARM loop.
Shot noise in the CARM, PRCL, MICH and SRCL (shown as ``Aux. DoF'')  are added
at each sensor,
and are mixed into the DARM loop 
through the non-diagonal interferometer (IFO) response (lower dashed arrow).
The feeedforward path in the simulation works between the DARM control signal
and control signal of the auxiliary DoFs as shown in the gray dashed trace on upper right.
``A'' is the actuator for DARM or Aux.}
\label{fig:control}
\end{center}
\end{figure*}

The fact that each sensor is sensitive to all DoFs
(in the other words, the sensing matrix is not completely diagonal),
leads to difficulties with interferometer sensing and control.
When the sensor signal at a given port is fed back to its target interferometer DoF,
some non-target-DoF signal is also fed back to that DoF due to these cross-couplings.

The DARM signal at the anti-symmetric ports (AS and OMC DC) can not be
cleanly separated from that of MICH because---from the point of view of the dark port---differential arm or Michelson motion are essentially equivalent (the gain at the AS port is much larger for DARM than for MICH because
DARM is enhanced by the arm cavities).
Therefore, there is always a MICH signal mixing into the DARM feedback signal,
even if the displacement in the MICH DoF is suppressed to the shot-noise level:
Because the optical gain of MICH is smaller than that of DARM, which is enhanced by the resonant arm cavity,
the shot noise limited sensitivity of MICH is worse than DARM.
As a consequence, the cross-coupled MICH signal increases the effective shot noise of the DARM sensor. 

As for second-order effects, the SRCL, PRCL and CARM signals are also mixed into DARM
through the MICH mixture path  \cite{somiya}.
The SRCL, PRCL, and CARM sensors have higher shot noise level than that of DARM,
because they are (in general) extracted at REFL or POP, both of which have higher light power
than the DARM port at the dark fringe.
Also, SRCL is almost always degenerate with the other DoFs and difficult
to extract independently because the finesse of SRC is much lower than those of the other DoFs.
Therefore, SRCL has a relatively low signal-to-noise ratio,
and the shot noise at the SRCL sensor appears in the displacement sensitivity
of DARM by way of the MICH mixture path.

Fig.~\ref{fig:control} shows a simplified diagram of the control loops.
For example, shot noise is added at each sensor into its respective loop,
and coupled into the DARM error signal.
Quantum noise shown in solid traces in Fig.~\ref{fig:DARMloop}
is by the following coupling process;
The shot noise in each sensor $ \vec{ V}_{\rm shot}$ propagates to each error signal as
\begin{eqnarray}
\vec{V}_{\rm errS} = (\mathbbm{1} - \mathbb{G}) ^{-1} \vec{ V}_{\rm shot}
\label{eq:qncoupling}
\end{eqnarray}
where $\mathbbm{1}$ is an identity matrix,
$\mathbb{G}$ is  the open-loop transfer function (in the frequency domain) 
of the five DoF loop including the servo filter shapes, $\mathbb{F}$,
actuator matrix, $\mathbb{A}$,
input matrix, $\mathbb{I}$, output matrix, $\mathbb{O}$
and the optical response as a matrix, $\mathbb{M}$.
Input and output matrices are the matrices to change either
from the sensor basis to the DoF basis, or,
from the DoF basis to the actuator basis, respectively
(matrix form of Table~\ref{tb:inout}, in the other words).
Here, $\mathbb{G, M, I}$ and $\mathbb{O}$ are $5\times 5$ matrices.
And $\mathbb{F}$ and $\mathbb{A}$ are $5\times 5$ diagonal matrices.
Note that the radiation pressure is also summed as the quantum noises in Fig.~\ref{fig:DARMloop}
in the both coupling processes.
The solid traces in Fig.~\ref{fig:DARMloop} is the DARM element of Eq.~\ref{eq:qncoupling}.

On the other hand, control noises,
the dashed traces in Fig.~\ref{fig:DARMloop}, are expressed as
\begin{eqnarray}
\vec{V} _{\rm errC} = (\mathbbm{1}-\mathbb{G}) ^{-1} \mathbb{IMAO} \vec{V}_{\rm ctrl}
\label{eq:ctrlcoupling}
\end{eqnarray}
where $\vec{V}_{\rm ctrl}$ is the control signal vector for the five DoFs.
Here, for the control signal, the shot noise
and radiation pressure noises are considered in $ \vec{V}_{\rm ctrl}$.

The calibrated displacement sensitivity $d \vec{L}$ is written as
\begin{eqnarray}
d\vec{L} = \mathbb{M}^{-1} (1+\mathbb{G}) \vec{V} _{\rm err}.
\label{eq:calb1}
\end{eqnarray}
As neither $\mathbb{G}$ nor $\mathbb{M}$ is diagonal, shot noise
at every sensor pollutes the DARM sensitivity.
The optical cross-coupling path is shown as gray dashed line on the bottom left in Fig.~\ref{fig:control}.

\subsection{Calibration}
To calibrate the DARM displacement, $dL_{\rm DARM}$ [m/$\sqrt{\rm Hz}$]
(the DARM element in Eq.~\ref{eq:calb1}),
into strain sensitivity, $h$ [1/$\sqrt{\rm Hz}$], down to low frequencies,
we have to take into account the suspension transfer function of the test masses
(or opto-mechanical transfer function when optical spring effect gives rise).
Treating the gravitational wave as tidal force, we have
\begin{eqnarray}
\label{eq:calb}
dL_{\rm DARM} = - \frac{dL}{dF} m \omega ^2 h L
\end{eqnarray}
where $L$ is the arm length, $\omega$ is the angular frequency of the signal,
$\frac{dL}{dF} $ is the (opto-) mechanical transfer function.
Note that Eq.~\ref{eq:calb} is valid where the gravitational wave length is
larger than $L$.
In our case of $L\sim4$\,km, this calibration method is valid from DC to 75\,kHz~\cite{Saulson}.

\subsection{Feedforward}
\label{sec:FF}
A feedforward technique is used to partially suppress the coupling
from the auxiliary DoF control signals into the DARM control signal.
As evidenced by the off-diagonal elements of the sensing matrix,
there are cross couplings from all the auxiliary DoFs into DARM.
In particular the SRCL signal cannot be extracted independently,
and it creates a large second-order coupling.
In our simulation,
the couplings from SRCL to DARM, and from MICH to DARM
are subtracted from the control signal by the feedforward path (dashed gray line in Fig.~\ref{fig:control})
so as to subtract the undesired contributions from SRCL and MICH above 20\,Hz.
The suppression ratios due to the feedforward at 20 Hz are
0.4 \% for MICH and 4 \% SRCL control noise.

This feedforward method was used in the first-generation GW detector Initial LIGO
and is planned to be implemented in aLIGO, as well.

\end{document}